\pdfoutput=1
\documentclass[useAMS,usenatbib,pdflscape]{mnras}
\usepackage[T1]{fontenc}
\usepackage{graphicx}
\usepackage{epstopdf}
\usepackage{amsmath}
\usepackage{amssymb}
\usepackage{rotating}
\usepackage{appendix}
\usepackage{subfigure}
\usepackage{stfloats}
\usepackage{subfigure}

\title[]{Time-dependent global simulations of a thin accretion disc: the effects of magnetically driven winds on thermal instability}

\author[Zhao et al.]{
Yu Zhao$^{1,2,3}$,
Xiao-Hong Yang$^1$\thanks{E-mail: yangxh@cqu.edu.cn (XH)},
Li Xue$^4$\thanks{E-mail: lixue@xmu.edu.cn (XL)},
and Shuang-Liang Li$^5$ \\
$^{1}$ Department of Physics and Chongqing Key Laboratory for Strongly Coupled Physics, Chongqing University, Chongqing 400044, China \\
$^{2}$ Key Laboratory for Particle Astrophysics, Institute of High Energy Physics, Chinese Academy of Sciences, \\
       19B Yuquan Road, Beijing 100049, China \\
$^{3}$ School of Physical Science, University of Chinese Academy of Sciences, 19A Yuquan Road, Beijing 100049, China\\
$^{4}$ Department of Astronomy, Xiamen University, Fujian, Xiamen 361005, China\\
$^{5}$ Key Laboratory for Research in Galaxies and Cosmology, Shanghai Astronomical Observatory, Chinese Academy of Sciences, \\
80 Nandan Road, Shanghai 200030, China \\
}

\begin{document}

\pagerange{\pageref{firstpage}--\pageref{lastpage}} \pubyear{20**}

\maketitle

\label{firstpage}
\def\LSUN{\rm L_{\odot}}
\def\MSUN{\rm M_{\odot}}
\def\RSUN{\rm R_{\odot}}
\def\MSUNYR{\rm M_{\odot}\,yr^{-1}}
\def\MSUNS{\rm M_{\odot}\,s^{-1}}
\def\MDOT{\dot{M}}

\begin{abstract}
According to the standard thin disc theory, it is predicted that the radiation-pressure-dominated inner region of a thin disc is thermally unstable, while observations suggest that it is common for a thin disc of more than 0.01 Eddington luminosity to be in a thermally stable state. Previous studies have suggested that magnetically driven winds have the potential to suppress instability. In this work, we implement one-dimensional global simulations of the thin accretion disc to study the effects of magnetically driven winds on thermal instability. The winds play a role in transferring the angular momentum of the disc and cooling the disc. When the mass outflow rate of winds is low, the important role of winds is to transfer the angular momentum and then shorten the outburst period.  When the winds have a high mass outflow rate, they can calm down the thermal instability. We also explore the parameter space of the magnetic field strength and the mass loading parameter.
\end{abstract}


\section{INTRODUCTION}
Observations on X-ray binaries have suggested that except for GRS 1915+105 and IGR J17091--3624, the high/soft state of X-ray binaries is quite stable in the luminosity ($L$) range of 0.01$--$0.5 Eddington luminosity ($L_{\text{Edd}}$) \citep{Belloni2000, Altamirano2011, Done2004}. This is inconsistent with the standard thin disc theory, which predicts that the radiation-pressure-dominated inner region of a thin disc is both thermally and viscously unstable when the accretion rate reaches $0.06 L_{\text{Edd}}$ \citep{Shakura1973, Piran1978}. For GRS 1915+105 and IGR J17091--3624, they are known to exhibit a limit-cycle light curve \citep{Belloni2000, Janiuk2002, Altamirano2011}, which is believed to arise from the thermal instability present in the thin disc \citep{Belloni1997, Szuszkiewicz2001, Li2007}. Although the time-scale of the limit-cycle light curve in active galactic nuclei (AGNs) is too long to be directly observed, the outbursts in some young radio galaxies appear to be triggered by the thermal instability \citep{Czerny2009, Wu2009}. Additionally, the periodic outbursts observed in the repeating changing-look (CL) AGNs are also thought to be a consequence of the thermal instability in the intermediate zone between an inner advection-dominated accretion flow and outer thin disc \citep{Sniegowska2020, Pan2021}. These studies at least indicate that it is common for the thin disc of $L \gtrsim0.01 L_{\text{Edd}}$  to be in a thermally stable state \citep{Gierlinski2004}. Regarding the limit-cycle behaviour, the intermittent outbursts in some young radio galaxies, and the periodic outbursts in the repeating CL AGNs, two possibilities can be considered: 1) there might be other physical processes for the outbursts, 2) the occurrence of thermal instability in the thin disc is contingent upon specific conditions.

The thermal instability of a thin disc is still an open issue. In theory, eliminating the thermal instability of thin discs is very necessary. The thermal instability arises from an imbalance between heating and cooling in the radiation-pressure-dominated inner region \citep{Shakura1976, Pringle1976}. Therefore, an in-depth investigation of the mechanisms responsible for heating and cooling is vital for understanding and resolving this issue. The heating processes in thin discs are believed as releasing the gravitational energy through the turbulence driven by the magnetorotational instability (MRI) \citep{Balbus1991}. In the standard thin disc theory, the $\alpha$-viscosity is employed to mimic the turbulence \citep{Shakura1973}. Some works have studied that when the viscous stress in the disc is  solely proportional to the gas pressure, the disc is stable \citep{Lightman1974}; when the viscous stress is proportional to the total pressure of gas and radiation, the disc is unstable \citet{Shakura1976}. This point has been confirmed by \citet{Ohsuga2006} through radiation hydrodynamical (RHD) simulations. Furthermore, radiation magnetohydrodynamical (RMHD) simulations conducted by \cite{Hirose2009} supported that the viscous stress is proportional to the total pressure. \citet{Jiang2013} have also found through the RMHD simulations that the disc is thermally unstable. In theory, for the disc to be stable, it is necessary to consider processes that cooling the disc or decreasing the relative importance of radiation pressure compared to the total pressure. For instance, \citet{Zheng2011} suggest that the magnetic pressure contributes to the vertical support force, reducing the relative importance of radiation pressure and thus stabilizing the disc. This finding has been confirmed by numerical simulations implemented by \citet{Sadowski2016}. \citet{Li2014} propose that the magnetically driven winds cool the disc, promoting stability. \citet{Jiang2016} find that the presence of an iron opacity "bump" can absorb heat in thin discs of AGNs, thereby maintaining disc stability.

Previous studies have suggested that magnetic pressure and magnetically driven flows play a important role in eliminating thermal instability \citep{Merloni2003, Zheng2011, Li2014, Habibi2019}. However, these studies primarily focused on analysing the instability and did not investigate the time-dependent evolution of the thin disc with poloidal and toroidal magnetic fields. The toroidal magnetic fields can support the thin disc by replacing radiation pressure \citep{Zheng2011}, while the poloidal magnetic fields can produce winds/outflows that cool the thin disc and transfer angular momentum \citep{Li2014}. In this work, we aim to simulate the time-dependent evolution of the thin disc with magnetic fields to study the effects of magnetically driven winds on thermal instability.

The paper is organized as follows. In section 2, we describe basic equations. In section 3, we present our results and related discussions. In section 4, we give conclusions and discussion.

\section{Basic equations}
In the standard thin disc around a black hole (BH), thermal instability may occur. Compared to the previous works on the time-dependent global simulations of thermal instability, our global simulations consider the effect of magnetic fields. The magnetic fields include the toroidal ($B_{\phi}$) and poloidal ($B_{\rm p}$) components. At the disc surface, the poloidal component is $\mathbf{B_{\rm p}}\equiv B_{r}\mathbf{e}_{r}+B_{z}\mathbf{e}_{\rm z}$ and ${B}_{\rm p}\equiv \sqrt{B_{r}^2+B_{z}^2}$, where $B_{r}$ is the radial component at the surface and $B_{\rm z}$ is the vertical component, respectively. In order to study the effect of the magnetocentrifugal-force-driven winds, the inclination angle of the field lines at the disc surface is requested to be less than 60$^{\rm o}$ with respect to the cold disc surface \citep{Blandford1982}. Therefore, $B_{\rm z}=\sqrt{3}B_{\rm r}$ at the disc surface is employed. In the accretion disc, the total pressure ($p$) at the mid-plane includes the gas pressure ($p_{\rm g}$), the radiation pressure ($p_{\rm r}$) and the magnetic pressure ($p_{\rm m}=\frac{B^2}{8 \pi}$), and reads
\begin{equation}
p=p_{\rm g}+p_{\rm r}+p_{\rm m}.
\end{equation}
Here, $p_{\rm g}=k\rho T/(\overline{\mu} m_{\rm H})$, where $k$, $m_{\rm H}$, $\overline{\mu}$, $\rho$, and $T$ are the Boltzmann constant, the hydrogen atom mass, the mean molecular weight, the density at mid-plane, and the temperature at mid-plane, respectively. We set $\overline{\mu}=0.62$. The radiation pressure is determined by
\begin{equation}
p_{\rm r}=\frac{F^-}{12c}(\tau_{\rm R}+\frac{2}{\sqrt{3} }),
\end{equation}
where $c$ is the light speed, $\tau_{\rm R}$ the Rosseland mean optical depths from the mid-plane to the disc surface, and $F^{-}$ is the radiative flux away from the disc surface, respectively. The radiative flux reads
\begin{equation}
F^-=\frac{24\sigma T^4}{\frac{3}{2}\tau_{\rm R}+\sqrt{3}+\frac{1}{\tau_{\rm p}} },
\end{equation}
where $\sigma$ is the Stefan--Boltzmann constant, and $\tau_{\rm p}$ the Plank mean optical depths from the mid-plane to the disc surface, respectively. Equations (2) and (3) are valid for both optically thick and thin regimes \citep{Szuszkiewicz1998}. When the accretion disc is extremely optically thick, i.e. $\tau_{\rm R}$ and $\tau_{\rm p}$ are much larger than unity, Equations (2) and (3) can reduce to the standard expressions in the standard thin disc. For $\tau_{\rm R}$ and $\tau_{\rm p}$, we follow \citet{Szuszkiewicz1998} and have
\begin{equation}
\tau_{\rm R}=0.34\Sigma[1+(6\times 10^{24}\rho T^{-3.5})],
\end{equation}
and
\begin{equation}
\tau_{\rm p}=\frac{1.24\times 10^{21}\Sigma\rho T^{-3.5}}{4\sigma}.
\end{equation}
Here, $\Sigma$ is the surface density of the accretion disc and reads $\Sigma=2 \rho H$, where $H$ is the half-thickness of the accretion disc.

The global evolution of the thin disc can be described using a set of hydrodynamical equations in cylindrical polar coordinates ($r$, $\phi$, $z$), where $r$ is the disc radius, $z$ is the symmetric axis of the disc, and $\phi$ is the azimuth, respectively. The mid-plane of the disc is the plane at $z=0$. For simplicity, we assume the disc to be axisymmetic (i.e. $\partial/\partial \phi\equiv0$) and vertically average the hydrodynamical equations. Therefore, the basic quantities of the disc properties, such as surface density ($\Sigma$), radial velocity ($v_{r}$), specific angular momentum ($l$), and temperature ($T$), are functions of time and radius. According to conservation of mass, the evolution of the surface density reads
\begin{equation}
\frac{\partial{\Sigma}}{\partial{t}}=-v_{r}\frac{\partial{\Sigma}}{\partial{r}}-\frac{\Sigma}{r}\frac{\partial}{\partial{r}}(rv_{r})-2\dot{m}_{\rm w},\label{mass}
\end{equation}
where $2\dot{m}_{\rm w}$ represents the mass-loss rate on unit area of disc surface due to the magnetocentrifugal-force-driven winds \citep{Pan2021} and is given by
\begin{equation}
\dot{m}_w=\frac{B_pB_z}{4\pi r \Omega }\mu, \label{mdot_w}
\end{equation}
where $\mu$ is the mass loading parameter and $\Omega$ is the angular velocity, respectively. The evolution of radial velocity reads
\begin{equation}
    \frac{\partial{v_r}}{\partial{t}}=-v_r\frac{\partial{v_r}}{\partial{r}}-\frac{1}{\rho}\frac{\partial{p}}{\partial{r}}+\frac{l^2-l^2_k}{r^3},\label{radialVelocity}
\end{equation}
where $l_{\rm k}$ is the Kepler angular momentum. In this work, we employ pseudo-Newton potential to mimic the general relativistic effects \citep{Paczysky1980} and then $l_k={\sqrt{GM_{BH}r^3}}/{(r-r_{\rm s})}$, where $G$ is the gravitational constant, $M_{BH}$ is the BH mass, and $r_{\rm s}\equiv\frac{2GM_{BH}}{c^2}$ is the Schwarzschild radius, respectively. The evolution of specific angular momentum reads
\begin{equation}
    \frac{\partial{l}}{\partial{t}}=-v_r\frac{\partial{l}}{\partial{r}}+\frac{\alpha}{r\Sigma}\frac{\partial}{\partial{r}}(r^3c_sH\Sigma\frac{\partial{\Omega}}{\partial{r}})-\frac{T_{\rm m}}{\Sigma}, \label{angularMoment}
\end{equation}
where $c_{\rm s}$ ($\equiv\sqrt{p/\rho}$), $\alpha$, and $T_{\rm m}$ are the local sound speed, the $\alpha$ parameter in the viscosity stress tensor, and the magnetic torque carried by the magnetic winds, respectively. $T_{\rm m}$ is given by
\begin{equation}
    T_m=\frac{B_p B_\varphi r }{2\pi}.\label{Tm}
\end{equation}
The evolution of temperature reads
\begin{equation}
\begin{aligned}
    \frac{\partial{T}}{\partial{t}}=-v_r\frac{\partial{T}}{\partial{r}}+\frac{T(1+\frac{1}{\beta_1})\frac{Q^+-F^-}{0.67pH}}{12-10.5\beta_2+\frac{4}{\beta_1}-\frac{3\beta_2}{\beta_1}}\\
    -\frac{T(4-3\beta_2+\frac{2}{\beta_1}-\frac{\beta_2}{\beta_1})(\frac{1}{r}\frac{\partial}{\partial{r}}(rv_{r})+2\frac{\dot{m}_w}{\Sigma}+\frac{v_z}{H})}{12-10.5\beta_2+\frac{4}{\beta_1}-\frac{3\beta_2}{\beta_1}},\label{Temperature}
\end{aligned}
\end{equation}
where we define $\beta_{1}=(p_{\rm g}+p_{\rm r})/p_{\rm m}$ and $\beta_{2}=p_{\rm g}/(p_{\rm g}+p_{\rm r})$ to measure the strength of magnetic fields and the ratio of gas pressure to the sum of gas and radiation pressure, respectively. Since the strength of magnetic fields cannot be determined inherently here, $\beta_1$ is a free parameter at $t=0$. $\beta_2$ depends on density and temperature.  $Q^{+}$ is the viscous heating rate and is given as
\begin{equation}
    Q^{+}=\alpha\Sigma c_{\rm s} H(r\frac{\partial{\Omega }}{\partial{r}})^2, \label{heating}
\end{equation}

Equations (\ref{mass}), (\ref{radialVelocity}), (\ref{angularMoment}), and (\ref{Temperature}) are consistent with conservation of mass, radial momentum, angular momentum and energy, respectively. To describe the evolution of $H$, we follow \citet{Li2007} and add two additional equations, as follows:
\begin{equation}
    \frac{\partial{H}}{\partial{t}}=-v_r\frac{\partial{H}}{\partial{r}}+v_z,\label{height}
\end{equation}

\begin{equation}
    \frac{\partial{v_z}}{\partial{t}}=-v_r\frac{\partial{v_z}}{\partial{r}}+6\frac{p}{\Sigma}-\frac{GM_{\rm BH}}{(r-r_{\rm s})^2}\frac{H}{r},\label{verticalVelocity}
\end{equation}
where $v_{\rm z}$ is the vertical velocity of the disc surface. When $v_{\rm z}>0$, the disc expands vertically. When $v_{\rm z}<0$, the disc shrinks vertically. The assumption of the vertical hydrostatic equilibrium is abandoned in this work. Therefore, Equations (\ref{mass}), (\ref{radialVelocity}), (\ref{angularMoment}), (\ref{Temperature}), (\ref{height}), and (\ref{verticalVelocity}) make up a set of equations to be solved. When magnetic fields are neglected, the equations are consistent with the equations in \citet{Li2007}.

The evolution of magnetic fields in the disc is essentially three-dimensional. The turbulence driven by MRI plays an important role not only in angular momentum transfer but also in the diffusion of magnetic fields \citep{Balbus1991, Guan2009, Fromang2009, Cao2013}. In our one-dimensional models, we cannot accurately describe MRI, while we can phenomenologically describe the advection and diffusion of magnetic fields. Phenomenologically, the change in the magnetic field strength is caused by the change in the surface density and the half-thickness. According to \cite{Li2014} 's assumption, we have
\begin{equation}
B_{t+\triangle t}=B_{t}(\frac{\Sigma_{t+\triangle t}}{\Sigma_{t}})^{\epsilon},\label{B}
\end{equation}
where the subscript $t$ and $t+\triangle t$ indicate that the values are physical quantities at time $t$ and $t+\triangle t$, respectively, and $\epsilon$ is a parameter depending on the diffusion and advection time-scales of magnetic fields. When $\epsilon=0$, corresponding to the case where the diffusion time-scale is far smaller than the advection time-scale, the strength of magnetic fields is a constant with the change of the surface density; When $\epsilon=1$, corresponding to the case where the advection time-scale is far smaller than the diffusion time-scale, all the magnetic flux ($\Phi=B/\Sigma$) is effectively advected inward and the flux is a constant \citep{Li2014}. Therefore, $0\leq \epsilon \leq 1$ is adopted. On the other hand, when the half-thickness ($H$) becomes larger (i.e. the disc expands vertically), the magnetic fields in the disc become weaker. This is supported by the MHD numerical simulations of a hot accretion flow \citep{Machida2006}. As in \citet{Zheng2011}, the toroidal component of magnetic fields is taken as a function of $H$, and given as
\begin{equation}
B_{\phi,t+\triangle t}=B_{\phi,t} (\frac{H_{t+\triangle t}}{H_{t}})^{-\gamma},\label{Bphi}
\end{equation}
where $\gamma$ is a free parameter. Equation (\ref{Bphi}) can keep $B_{\phi}H^{\gamma}$ constant with time. It is assumed that $\gamma \geq 1$. This implies that as the disc expands vertically, the toroidal magnetic field becomes weaker. When $\gamma =1$, the toroidal magnetic flux per unit radius is conserved.

\begin{table*}
    \begin{center}
    \caption[]{Parameters of models}
    \begin{tabular}{ccccccccccc}
    \hline\noalign{\smallskip} \hline\noalign{\smallskip}
    Models &$\frac{B_{\phi,0}}{B_{p,0}}$&$\beta_{1,0}$& $\mu$ &Outburst & Period(s) & $\frac{L_{max}-L_{min}}{L_{min}}$\\
    \hline\noalign{\smallskip}
    M1  &$\sim$ & $\infty$ & $0$ & Yes & $648$ & $235$  \\

    M2  &0.1& $100$ & $0.1$ & Yes & $535$ & $212$ \\
    M3  &0.1& $100$ & $1$ & Yes & $\infty$  & -- \\
    M4  &0.1& $100$ & $1.5$ & No & $\infty$ & --\\
    M5  &0.1& $100$ & $2$ & No & $\infty$  & --\\

    M6  &0.1& $200$ & $0.1$ & Yes & $542$  & $214$\\
    M7  &0.1& $200$ & $2$ & Yes & $\infty$  & --\\
    M8  &0.1& $200$ & $3$ & No & $\infty$  & --\\

    M9  &0.1& $300$ & $0.1$ & Yes & $546$  & $213$\\
    M10  &0.1& $300$ & $3$ & Yes & $\infty$  & --\\
    M11  &0.1& $300$ & $4$ & No & $\infty$  & --\\

    M12  &0.1& $400$ & $0.1$ & Yes & $552$  & 214\\
    M13  &0.1& $400$ & $4$ & Yes & $\infty$  & --\\
    M14  &0.1& $400$ & $5$ & No & $\infty$  & --\\

    M15  &0.1& $500$ & $0.1$ & Yes & $554$  & $215$\\
    M16  &0.1& $500$ & $5$ & Yes & $\infty$  & --\\
    M17  &0.1& $500$ & $6$ & No & $\infty$  & --\\

    M18  &0.1& $600$ & $0.1$ & Yes & $555$  & $215$\\
    M19  &0.1& $600$ & $6$ & Yes & $\infty$  & --\\
    M20  &0.1& $600$ & $7$ & No & $\infty$  & --\\

    M21  &0.1& $700$ & $0.1$ & Yes & $557$  & $215$\\
    M22  &0.1& $700$ & $7$ & Yes & $\infty$  & --\\
    M23  &0.1& $700$ & $8$ & No & $\infty$  & --\\

    M24  &0.1& $800$ & $0.1$ & Yes & $560$  & $215$\\
    M25  &0.1& $800$ & $8$ & Yes & $\infty$  & --\\
    M26  &0.1& $800$ & $9$ & No & $\infty$  & --\\

    M27  &0.1& $900$ & $0.1$ & Yes & $560$  & $216$\\
    M28  &0.1& $900$ & $9$ & Yes & $\infty$  & --\\
    M29  &0.1& $900$ & $10$ & No & $\infty$  & --\\

    M30  &0.1& $1000$ & $0.1$ &Yes& $561$ & $216$\\
    M31  &0.1& $1000$ & $10$ & Yes & $\infty$  & --\\
    M32  &0.1& $1000$ & $11$ & No & $\infty$  & --\\
    M33  &0.1& $1000$ & $15$ & No & $\infty$  & --\\

    M34  &10& $100$ & $0.1$ & Yes & $528$  & $212$\\
    M35  &10& $100$ & $5.0$ & Yes & $497$  & $201$\\
    M36  &10& $100$ & $150.0$ & No & $\infty$ & --\\

    M37  &10& $200$ & $0.1$ & Yes & $515$ & $201$\\
    M38  &10& $300$ & $0.1$ & Yes & $539$ & $202$\\
    M39  &10& $400$ & $0.1$ & Yes & $530$ & $202$\\
    M40  &10& $500$ & $0.1$ & Yes & $543$ & $203$\\
    M41  &10& $600$ & $0.1$ & Yes & $535$ & $203$\\
    M42  &10& $700$ & $0.1$ & Yes & $534$ & $202$\\
    M43  &10& $800$ & $0.1$ & Yes & $533$ & $200$\\
    M44  &10& $900$ & $0.1$ & Yes & $541$ & $203$\\
    M45  &10& $1000$ & $0.1$ & Yes & $542$ & $205$\\

   \hline\noalign{\smallskip}
    \hline\noalign{\smallskip}
    \end{tabular}
    \end{center}
    \begin{list}{}
    \item\scriptsize{\textit{Note}. Column (1): the model number; Column (2): the ratio of initial $B_{\phi,0}$ and $B_{p,0}$; Column (3): the initial value ($\beta_{1,0}$) of the parameter $\beta_1$; Column (4): the mass loading parameter ($\mu$); Column (5): "Yes" means that the model is unstable and then the outburst occurs; "No" means that the model is thermally stable and then the outburst does not occur; Column (6): the period of the limit-cycle behaviour; Column (7): luminosity variation amplitude. $\beta_{1,0}$ reflects the initial strength of the magnetic field. Smaller $\beta_{1,0}$ means a stronger initial magnetic field effect. In model M1, $\beta_{1,0}$ equals infinity, which means that the large-scale magnetic fields are not included in model M1.}
    \end{list}
\end{table*}

\begin{figure*}
\scalebox{0.3}[0.3]{\rotatebox{0}{\includegraphics{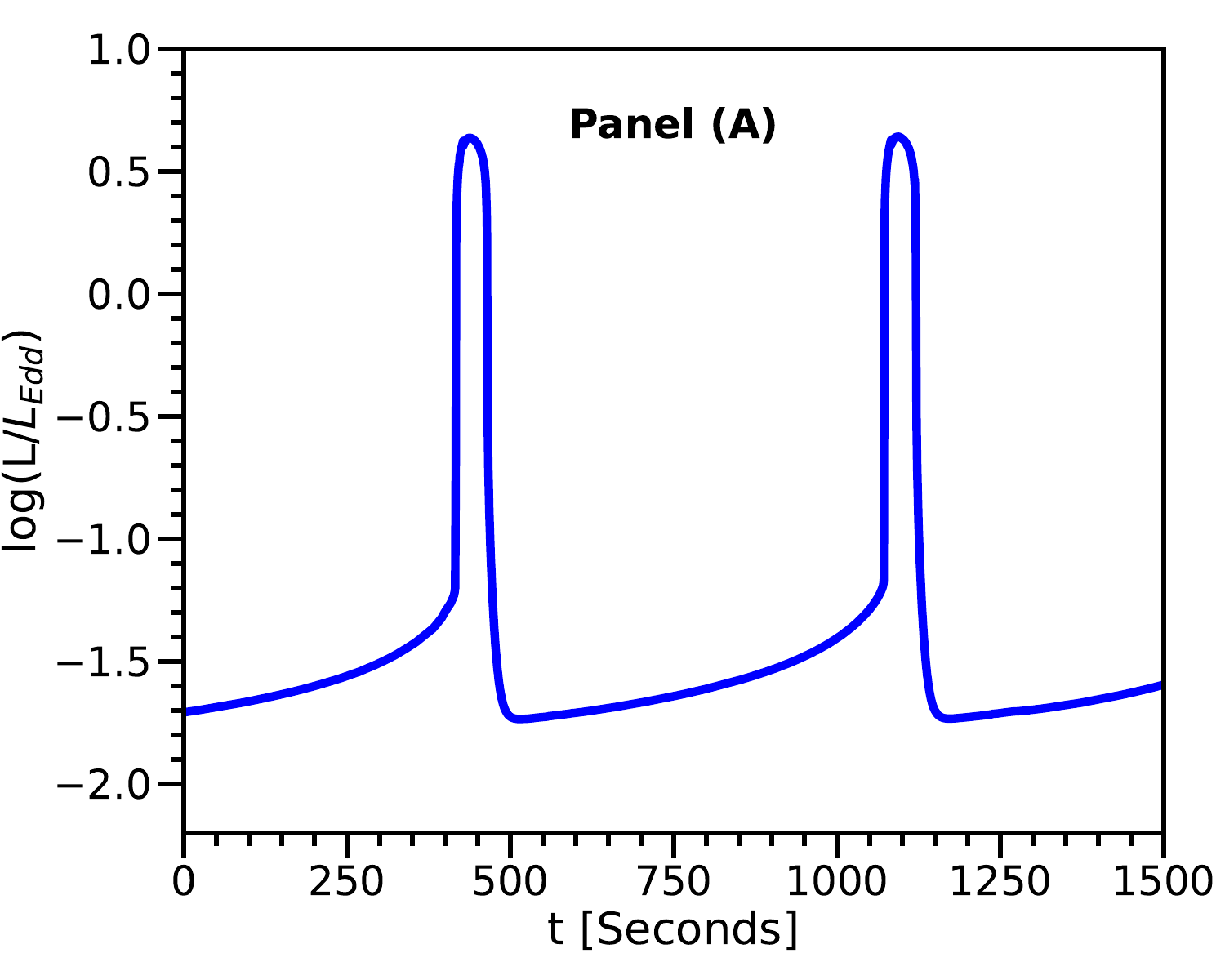}}}
\scalebox{0.3}[0.3]{\rotatebox{0}{\includegraphics{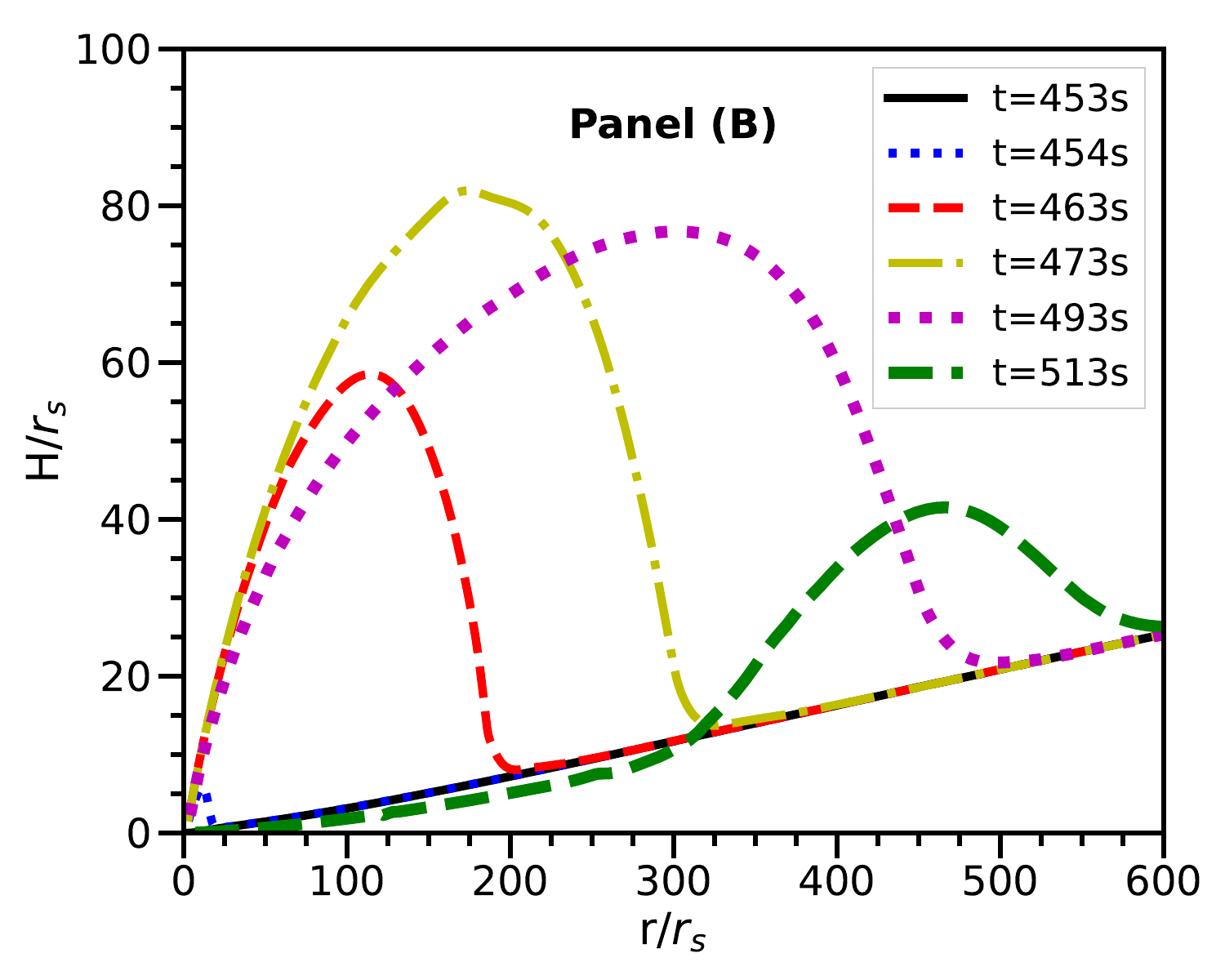}}} \\
\scalebox{0.3}[0.3]{\rotatebox{0}{\includegraphics{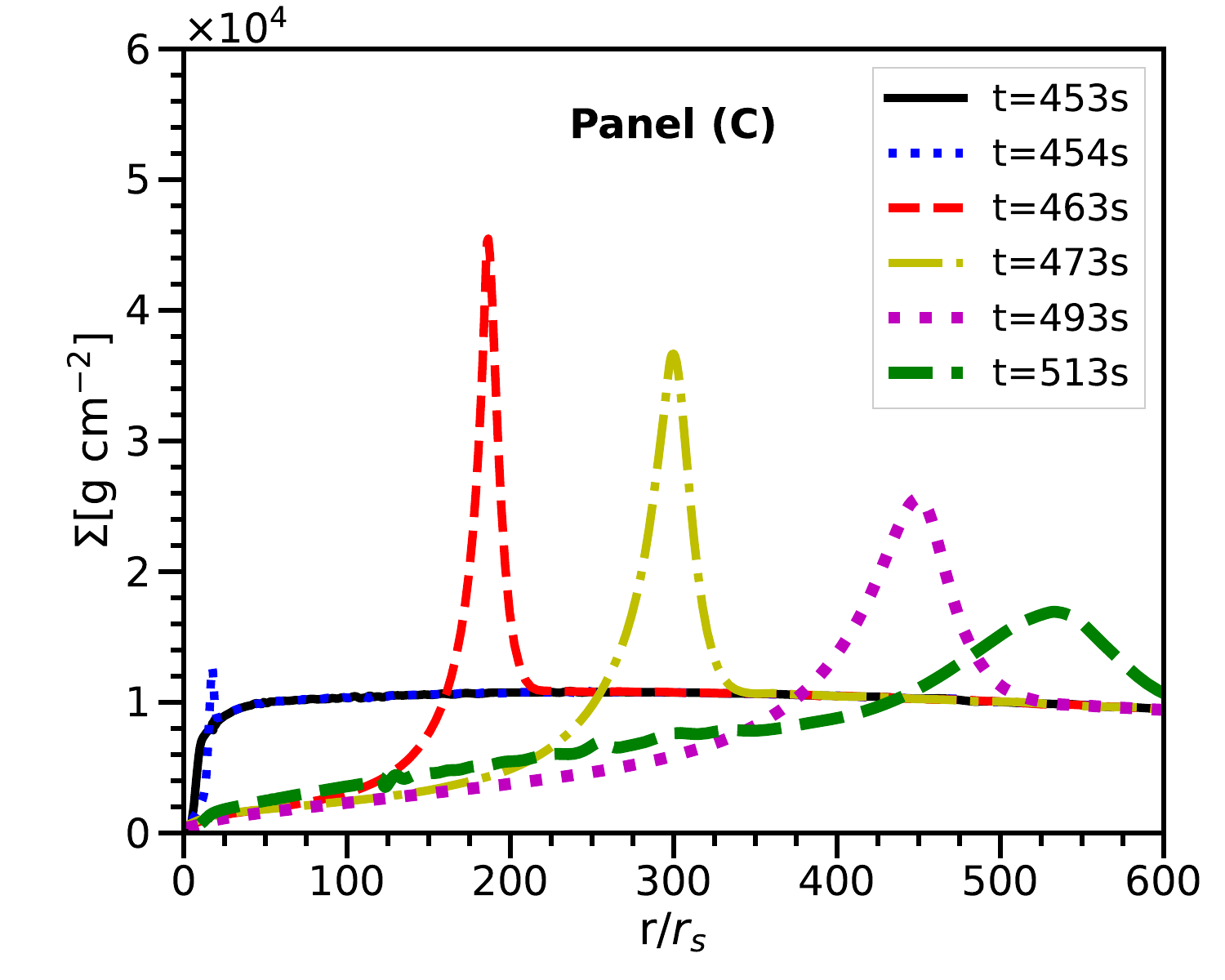}}}
\scalebox{0.3}[0.3]{\rotatebox{0}{\includegraphics{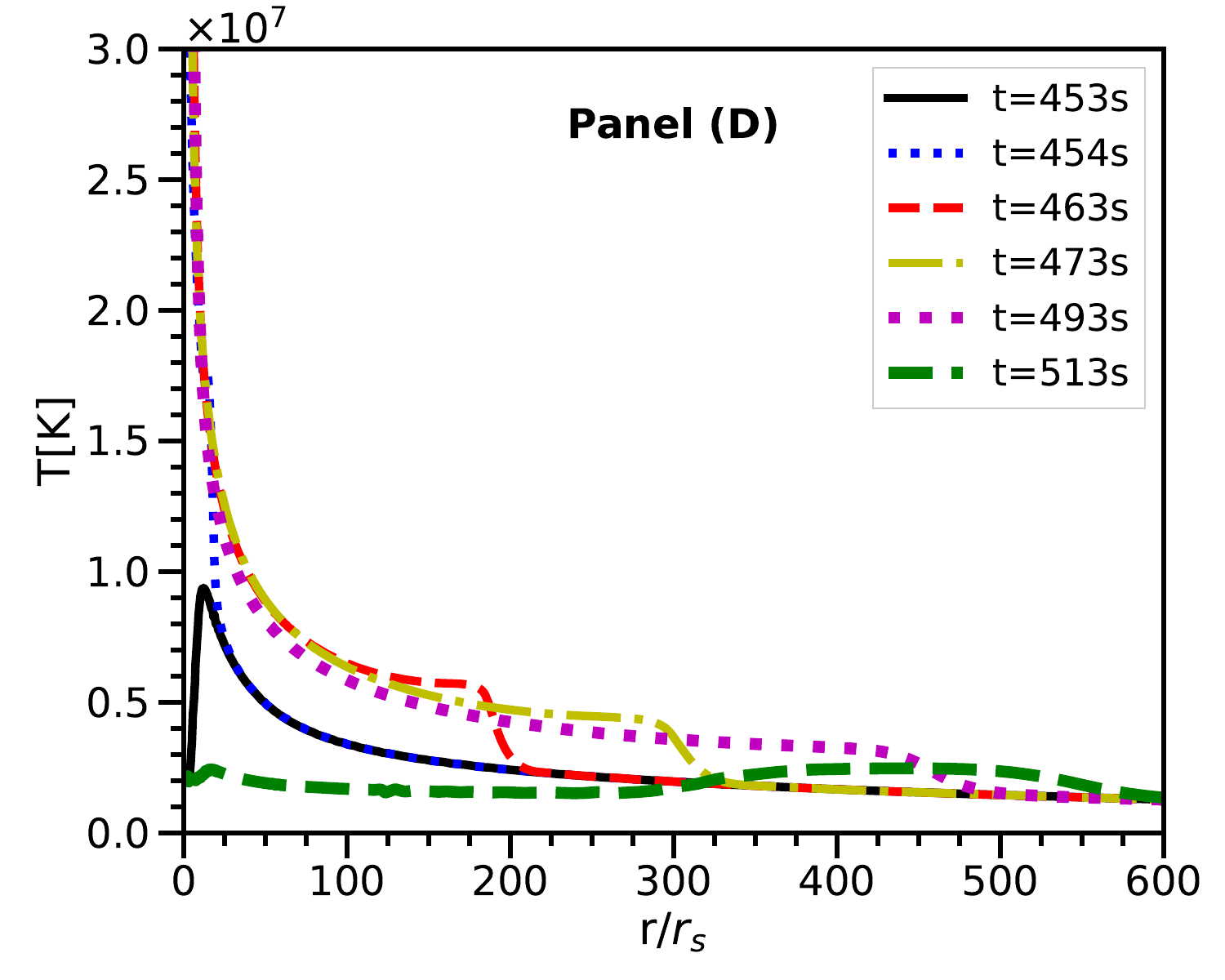}}}
\ \centering \caption{Evolutions of the disc luminosity [panel (A)], half height [panel (B)], surface density [panel (C)], and temperature [panel (D)] in model M1 without magnetic fields.}
\label{Fig1}
\end{figure*}

\begin{figure*}
\scalebox{0.3}[0.3]{\rotatebox{0}{\includegraphics{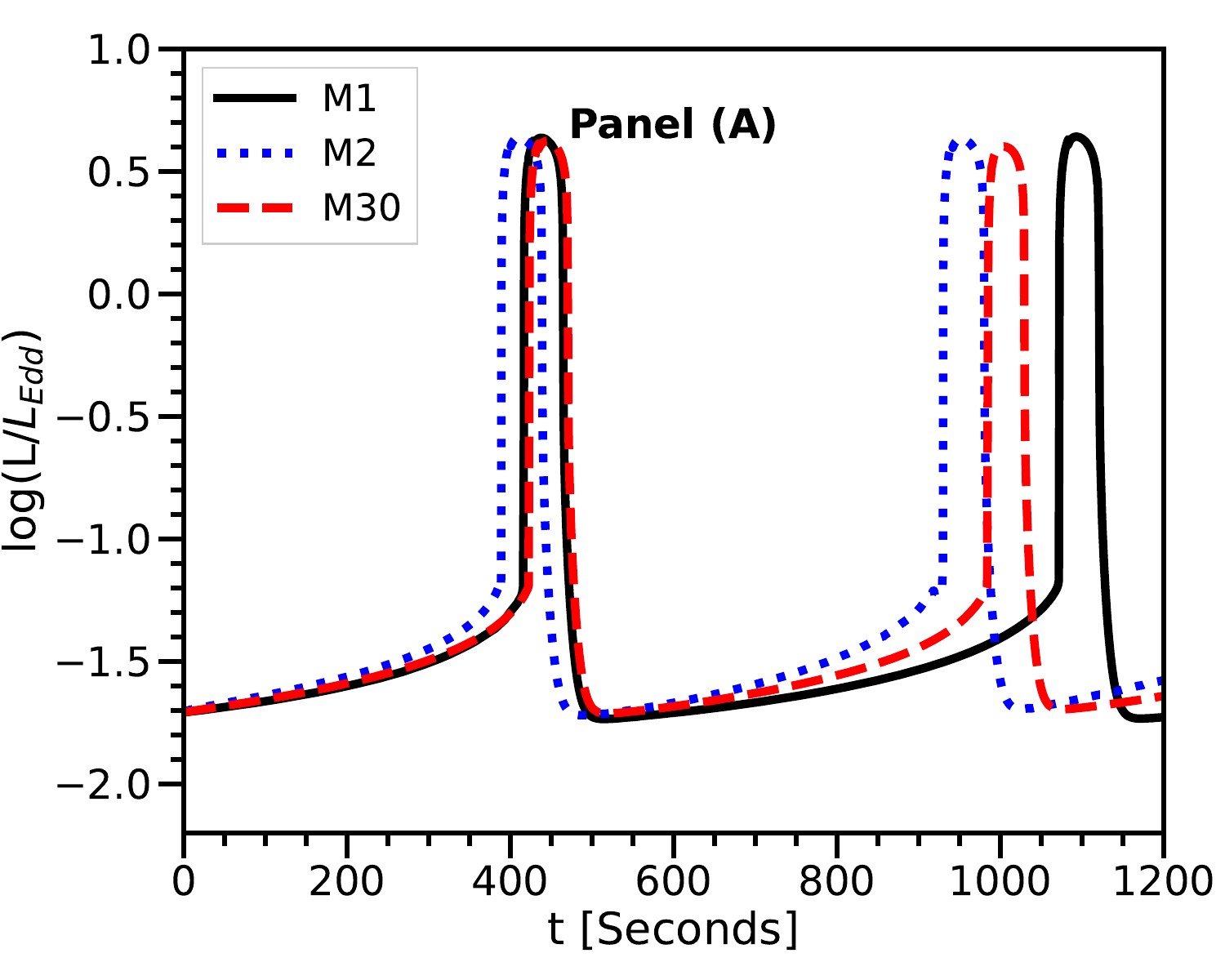}}}
\scalebox{0.3}[0.3]{\rotatebox{0}{\includegraphics{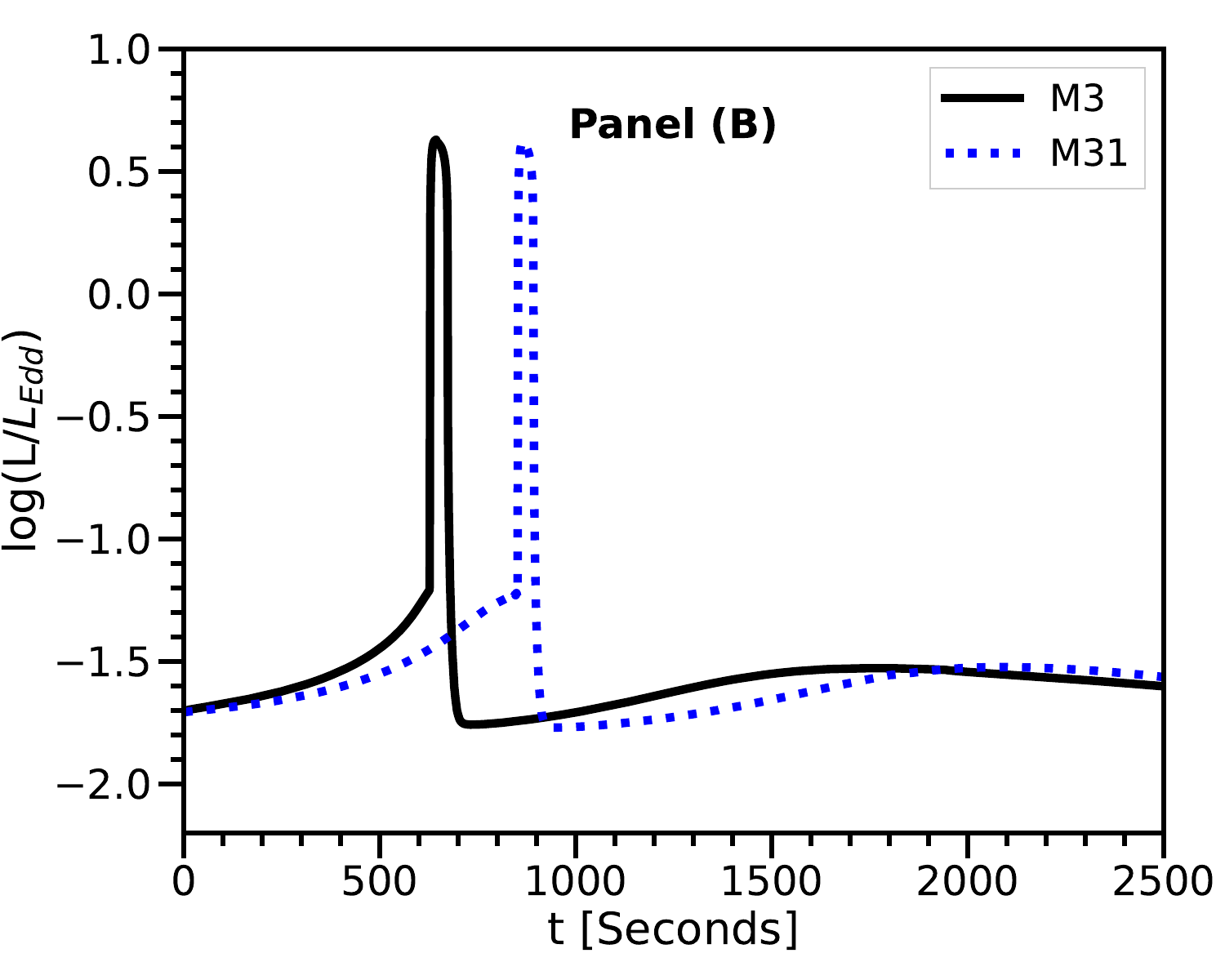}}}
\scalebox{0.3}[0.3]{\rotatebox{0}{\includegraphics{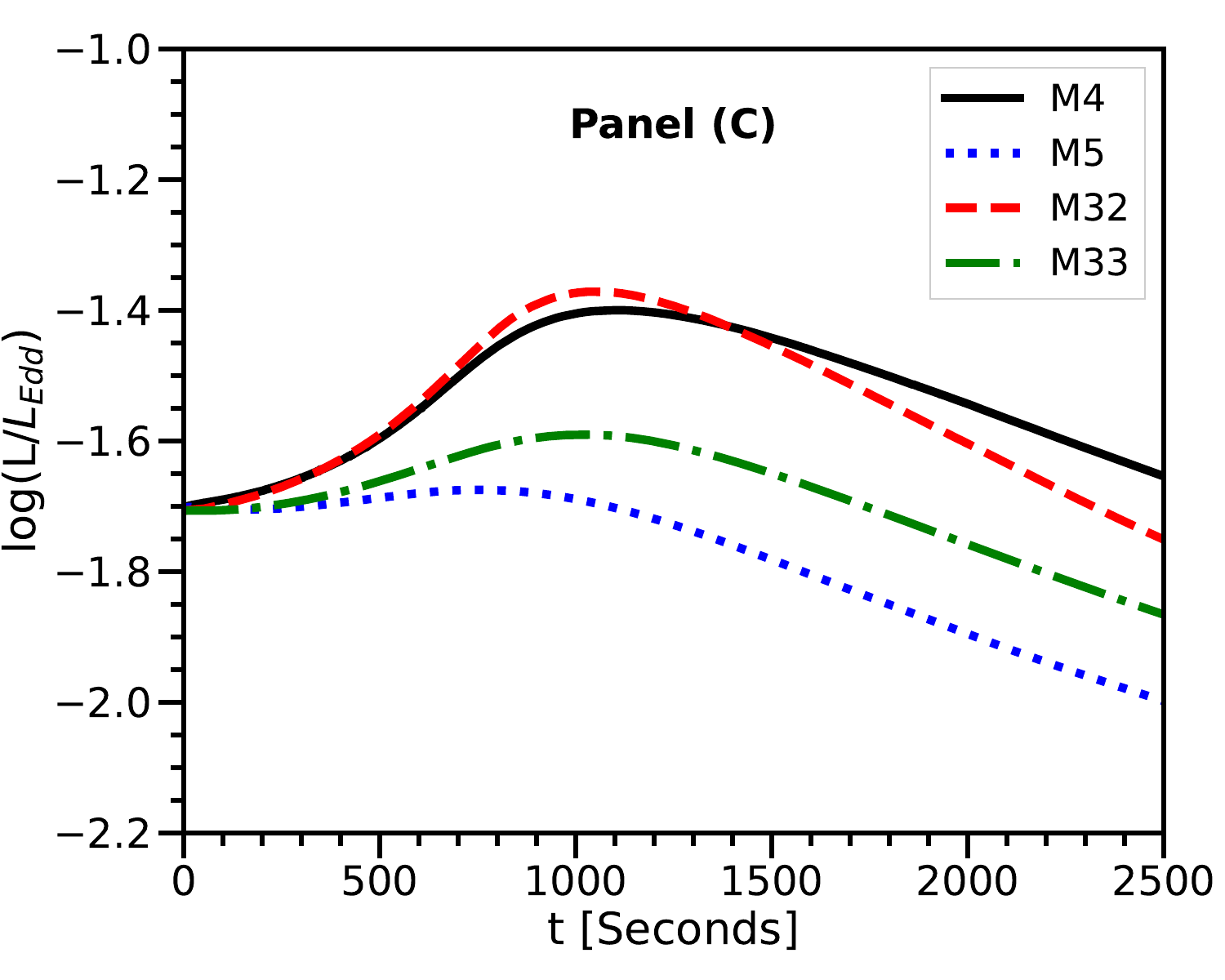}}}
\ \centering \caption{Evolutions of luminosity in the models M1--M5 and M30--M33 with the magnetic field structure of $B_{\phi,0}=0.1B_{p,0}$. Panel (A) shows that the three models exhibit the limit-cycle behaviour. Panel (B) shows that the two models experience an outburst and then calm down the next outburst. Panel (C) shows that the four models can calm down the outburst triggered by the thermal instability.}
\label{Fig2}
\end{figure*}

\section{Numerical methods and initial condition}
The dynamical quantities of the accretion disc, $\Sigma$,  $v_{r}$, $l$, $T$, $H$, and $v_{z}$, are obtained by solving a set of dynamical equations, which consist of the Equations (\ref{mass}), (\ref{radialVelocity}), (\ref{angularMoment}), (\ref{Temperature}), (\ref{height}), and (\ref{verticalVelocity}). The evolution of magnetic fields can be obtained from Equations (\ref{B}) and (\ref{Bphi}). For the time integration, a third-order total variation diminishing Runge-Kutta scheme is adopted to integrate the first two time steps only in the initial time ($t=0$) and then the third-order backward differentiation explicit scheme is used for the remaining computations to speed up the calculation. The standard Chebyshev pseudo-spectral method is implemented in discretizing the spatial difference. Computational domain is set to be between $2r_{\rm s}$ and $10^4 r_{\rm s}$. We have modified the code in \cite{Li2007} to include the effect of magnetic fields and implement the present simulations. More details on numerical methods are referred in \cite{Li2007}.

Our model parameters include $M_{\rm BH}$, the accretion rate at the outer boundary ($\dot{M}$), $\alpha$, $\mu$ in Equation (\ref{mdot_w}), $\epsilon$ in Equation (\ref{B}), $\gamma$ in Equation (\ref{Bphi}), and $\beta_{1,0}$, where $\beta_{1,0}$ is the initial value of $\beta_1$. In the present simulations, we set $M_{BH}=10M_\odot$, $\dot{M}=0.1\dot{M}_{\rm Edd}$ ($\dot{M}_{\rm Edd}\equiv 10 L_{\rm Edd}/c^2$  is the Eddington accretion rate), $\alpha=0.1$, $\epsilon=0.4$, and $\gamma=1$ respectively. Then, $\mu$ and $\beta_{1,0}$ become two main free parameters in our models. A transonic steady thin disc around a BH with $0.1\dot{M}_{\rm Edd}$  is taken as our initial condition. $\beta_{1,0}$ determines the strength of initial magnetic field. Moreover, we calculate two cases of $B_{\phi,0}=0.1B_{p,0}$ and $B_{\phi,0}=10B_{p,0}$, which represents the two cases where the polar and toroidal component dominate, respectively. For the initial poloidal component, we assume $B_{z,0}=\sqrt{3}B_{r,0}$. In this way, we can determine the initial structure of magnetic fields. Table 1 gives the basic parameters of our models.

\section{RESULTS}
As the initial model for numerical simulations, we employ a transonic thin disc without magnetic fields \citep{Abramowicz1988, Matsumoto1984}. We firstly calculate the evolution of the thin disc without magnetic fields and then calculate the evolution of the thin disc with weak magnetic fields.

\subsection{Evolution of the thin disc without magnetic fields}
In model M1, magnetic fields are ignored. Fig. \ref{Fig1} shows the evolutions of the luminosity ($L$), the half-height ($H$), the surface density ($\Sigma$), and the temperature ($T$), respectively.
This model exhibits the limit-cycle behaviour, which is consistent with \citet{Li2007}. All physical quantities evolve periodically, with one cycle lasting for 680s. When the disc evolves to an approximate thin disc in a low-luminosity state, we set $t=0$. The material is slowly accreted by a BH from the outer zone to the inner zone with time, making the surface density and temperature in the inner zone increase. This increase in radiation pressure ($\propto T^4$) triggers thermal instability in the inner zone. At around $452$s, the temperature rises rapidly, the disc height significantly expands, and the density peak drifts outward, carrying a large amount of material moving outward. The disc remains in a high-luminosity state for about 40s.

\subsection{Effects of magnetically driven winds}
In models M2--M33, the magnetic fields are dominated by a poloidal magnetic field and we set $B_{\phi,0}=0.1B_{p,0}$. In models M34--M45, the magnetic fields are dominated by a toroidal magnetic field and we set $B_{\phi,0}=10B_{p,0}$. The large-scale poloidal magnetic field can produce winds, and the mass outflow rate is described by Equation \ref{mdot_w}, which indicates that a stronger poloidal field can produce more powerful magnetic winds, and larger $\mu$ enhances the mass outflow rate of winds. To vary the initial strength of the magnetic fields in our models, we adjust the value of $\beta_{1,0}$.

In our simulations, we initially superimpose a weak magnetic field on the initial model. Superimposing the weak magnetic field is to avoid disrupting the structure of the initial model caused by the superimposed magnetic field. Moreover, when we superimpose a stronger magnetic field, our calculation becomes unstable. Therefore, $\beta_{1,0}$ is taken from 100 to 1000.

Fig. \ref{Fig2} shows the evolution of luminosity in some typical models. Panel (A) displays three models that exhibit limit-cycle behaviour, while panel (C) shows four models that do not display any outburst. In panel (B), two models experience an outburst and then keep calming down. This indicates that these models are in a transitional state from being completely unstable to completely stable. We think that these models are ultimately stable. Generally, when considering the same $\beta_{1,0}$, a larger value of $\mu$ is needed to make the disc stable. For example, the thin disc remains stable and does not transition to a high luminous state when $\mu$ exceeds 1.5 for $\beta_{1,0}=100$, and $\mu$ exceeds 10 for $\beta_{1,0}=1000$. If $\mu$ is small enough to neglect the effect of magnetic winds, the disc is still thermally unstable and exhibits the limit-cycle behaviour. In panel (A), the limit-cycle behaviour is displayed in models M1, M2, and M30. For the model M2 with $\beta_{1,0}=100$ and $\mu=0.1$, the model M30 with $\beta_{1,0}=1000$ and $\mu=1$, the limit-cycle behaviour takes place. Compared to model M1 without magnetic fields, the limit-cycle periods in these two models are shorter. Due to the weaker magnetic fields in model M30 compared to model M2, the limit-cycle period in model M30 is slightly longer than that in M2.

\begin{figure}
    \scalebox{0.35}[0.35]{\rotatebox{0}{\includegraphics{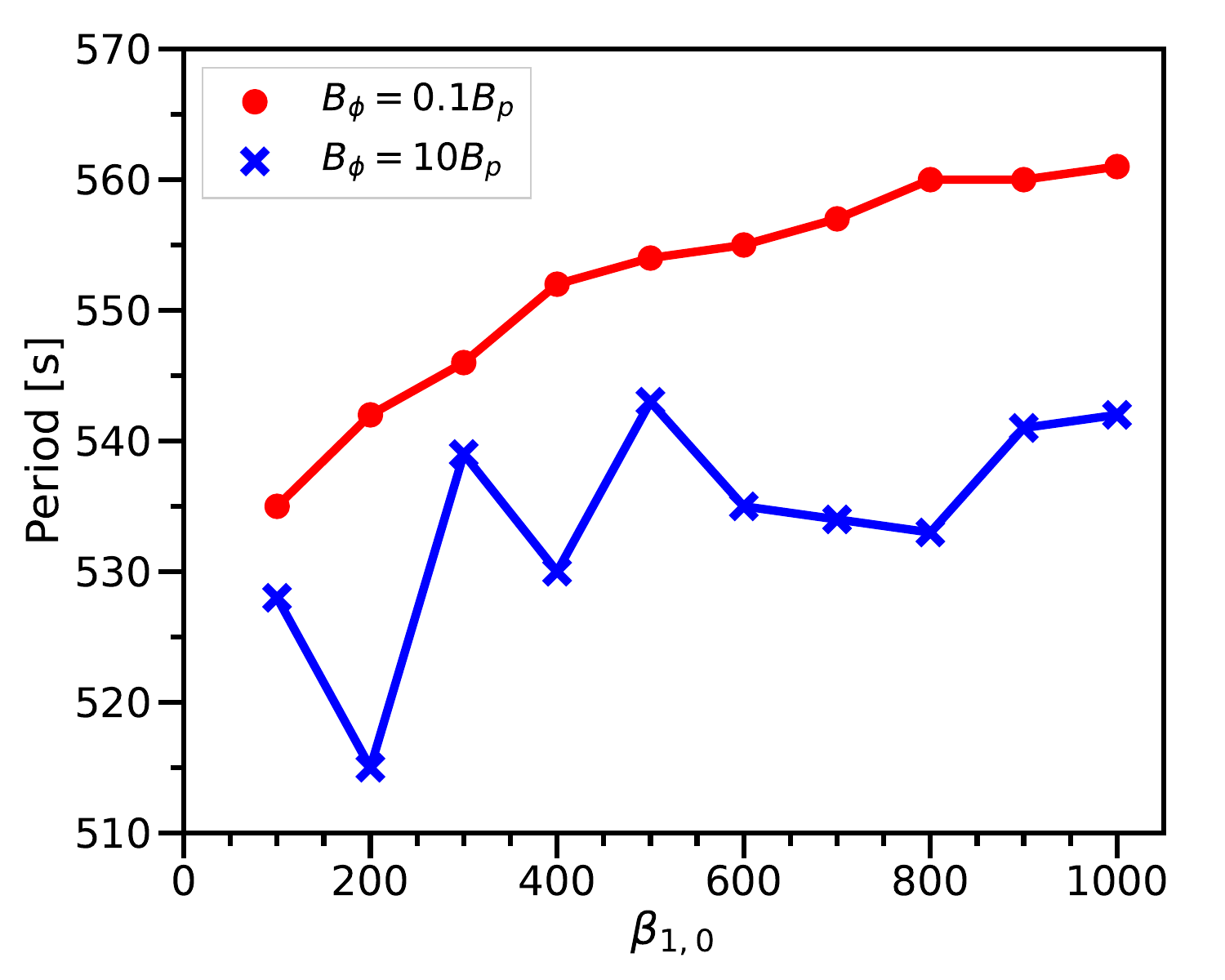}}}
    \ \centering \caption{The dependence of the limit-cycle period on the parameter $\beta_{1,0}$. The red line corresponds to the models of $B_{\phi}=0.1B_{\rm p}$, while the blue line corresponds to the models of $B_{\phi}=10B_{\rm p}$.}
    \label{Fig3}
\end{figure}

Fig. \ref{Fig3} further shows the dependence of the limit-cycle period on the parameter $\beta_{1,0}$: the models with the stronger magnetic fields have slightly shorter periods. In particular, the limit-cycle period in model M2 has decreased by about 20\% compared to model M1, as shown in Table 1. The limit-cycle period in the $B_{\phi}=10B_{p}$ case is shorter than that in the $B_{\phi}=0.1B_{p}$ case due to the stronger magnetic torque produced by the magnetic field structure in $B_{\phi}=10B_{p}$ case (Equation \ref{Tm}). The variation of period agrees with that in \citet{Pan2021} qualitatively, who numerically studied a local radiation-pressure-dominated thin disc with a poloidal magnetic field for understanding repeating CL AGNs. The limit-cycle period is reduced by the poloidal magnetic field because the magnetically-driven winds take away the angular momentum of the disc and then reduce the time interval for the material to fall back to the outburst region.

\begin{figure}
    \scalebox{0.35}[0.35]{\rotatebox{0}{\includegraphics{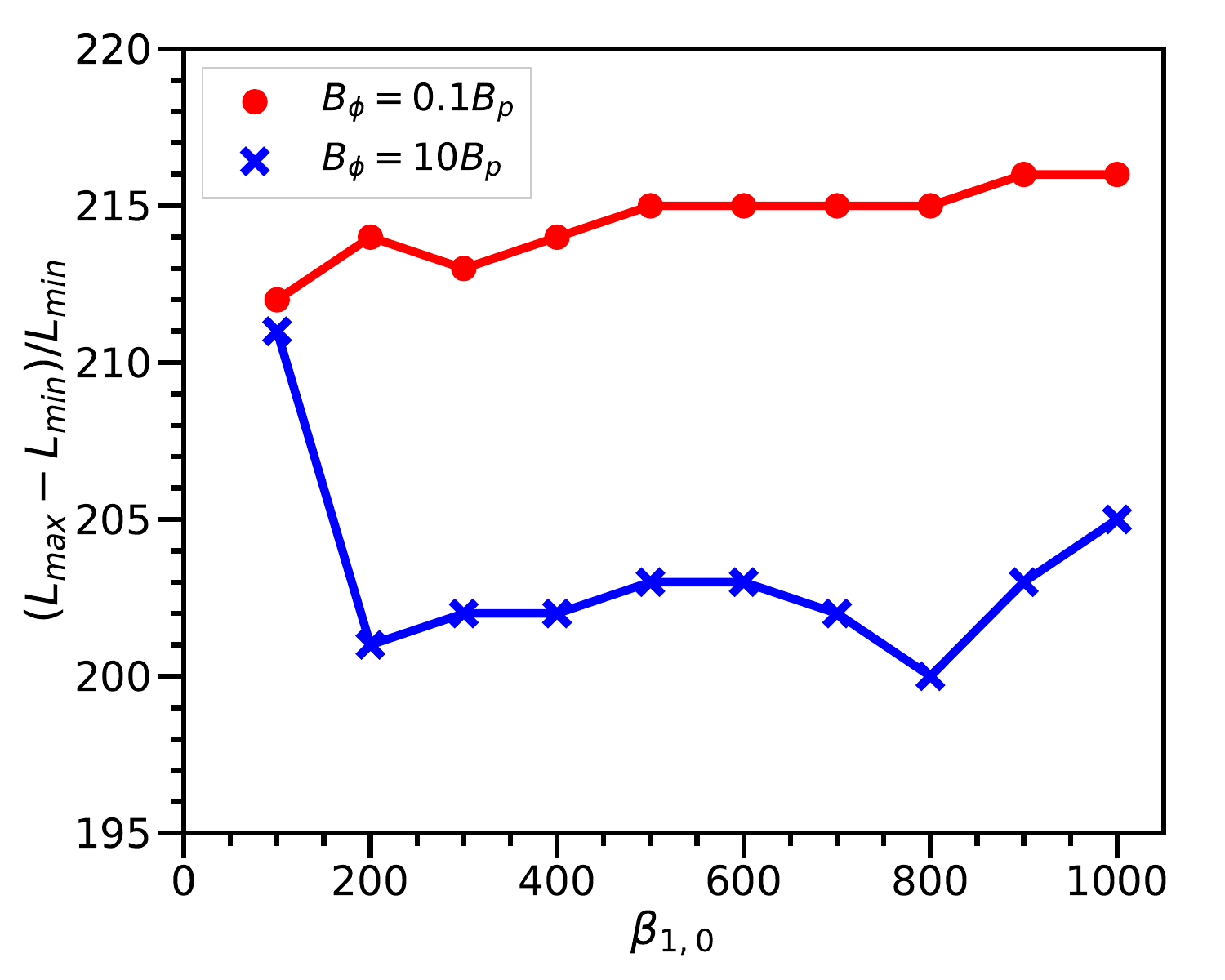}}}
    \ \centering \caption{The dependence of the luminosity variation amplitude on the parameter $\beta_{1,0}$. The red line corresponds to the models of $B_{\phi}=0.1B_{\rm p}$, while the blue line corresponds to the models of $B_{\phi}=10B_{\rm p}$.}
    \label{Fig4}
\end{figure}

Fig. \ref{Fig4} shows the dependence of the luminosity variation amplitude on the parameter $\beta_{1,0}$. In our models, the highest luminosity is almost the same for different $\mu$ and $\beta_{1,0}$, and the luminosity variation also is not significant. In the $B_{\phi}=10B_{p}$ case, the luminosity variation amplitude is slightly smaller than that in $B_{\phi}=0.1B_{p}$ case. However, in \citet{Pan2021}, the highest luminosity and variation amplitude are different for different $\mu$ and $\beta_{1,0}$. This is because our calculations are based on global simulations instead of the local area in thin disc, called the transition zone in \citet{Pan2021}. In our simulations, when an outburst takes place in the disc, the height and temperature of the inner region increase significantly, while the surface density decreases by outflow. This leads to an extremely weak magnetic field in the inner region (see Equation \ref{B} and Equation \ref{Bphi}). Consequently, the temperature and area of the hottest inner region are not significantly affected by the magnetic field during the outburst. Therefore, the maximum and minimum luminosity is almost the same in our models with different parameters.

For the two models (M3 and M31) shown in panel (B) of Fig. \ref{Fig2}, the initial introduction of magnetic fields does not immediately have a significant effect on cooling. Therefore, the disc firstly experiences an outburst, and at the end of the outburst, the disc's height rapidly shrinks and strengthens the magnetic fields to produce strong winds, so that the disc is effectively cooled by the winds over time and finally becomes thermally stable like the models [shown in panel (C) of Fig. \ref{Fig2}] with larger $\mu$.

For the models in a thermally stable state [panel (C) of Fig. \ref{Fig2}], the disc luminosity gradually increases, reaches a maximum value, and then gradually decreases. As an example, we show the evolution of the disc structures of model M4 in Fig. \ref{Fig5}. For the other thermally stable models, their disc structures are similar in evolution. As shown in Fig. \ref{Fig5}, during the first 1000s of disc accretion, a large amount of material is carried to move into inside $\sim200r_{\rm g}$ [as shown in panel (A)],  the disc height and temperature both increase, shown in panels (B) and (C).

However, in model M4, the increase of the disc temperature is significantly suppressed compared to model M1. With the accumulation of material inside $\sim200r_{\rm g}$, the magnetic fields in the inner region become stronger (as shown in panel (D)). Panel (E) indicates that the poloidal field becomes stronger than the initial fields. Panel (F) indicates that the toroidal field within $\sim15r_{\rm s}$ becomes stronger, while the toroidal field beyond $\sim15r_{\rm s}$ becomes weaker. According to Equations (15) and (16), the change of the toroidal field is determined by the surface density and disc height. Within $\sim15r_{\rm s}$, the toroidal field enhancement is dominated by the surface density increase. Beyond $\sim15r_{\rm s}$, the toroidal field weakening is dominated by the disc height increase. For the poloidal field, its strength change is only determined by the surface density. Therefore, the magnetic field enhancement is caused by the poloidal field enhancement. This results in the magnetically driven winds becoming so powerful that a significant portion of thermal energy is dissipated. Consequently, the disc temperature is constrained from further increasing, so the thermal instability is suppressed in model M4. After about 1000s, more and more material is lost through the magnetic driven winds, resulting a reduction in the surface density and subsequent weakening of the magnetic fields. Simultaneously, the disc temperature and luminosity decrease gradually. This is attributed to the diminishing viscous heating rate, which is proportional to the surface density (see Equation \ref{heating}).

Fig. \ref{Fig6} shows the boundary between stable and unstable areas in the $\mu$--$\beta_{1,0}$ parameter space for the $B_{\phi,0}=0.1 B_{p,0}$ case. The area above the line is stable for models, while the areas below the line are unstable. For the same strength of magnetic fields, increasing the mass loading parameter ($\mu$) is helpful to suppress thermal instability. If the mass loading parameter is small enough, the disc becomes thermally unstable and exhibits the limit-cycle behaviour. In this case, the role of a poloidal magnetic field is to transfer the angular momentum and then change the period of the limit-cycle behaviour. In the $B_{\phi,0}=10B_{p,0}$ case, the $\mu$ value on the boundary is 100 times higher than that in the $B_{\phi,0}=0.1 B_{p,0}$ case. As an example, model M36 becomes thermally stable when the $\mu$ value is 100 times higher than that in model M4.

\section{CONCLUSION AND DISCUSSION}

\begin{figure*}
    \scalebox{0.3}[0.3]{\rotatebox{0}{\includegraphics{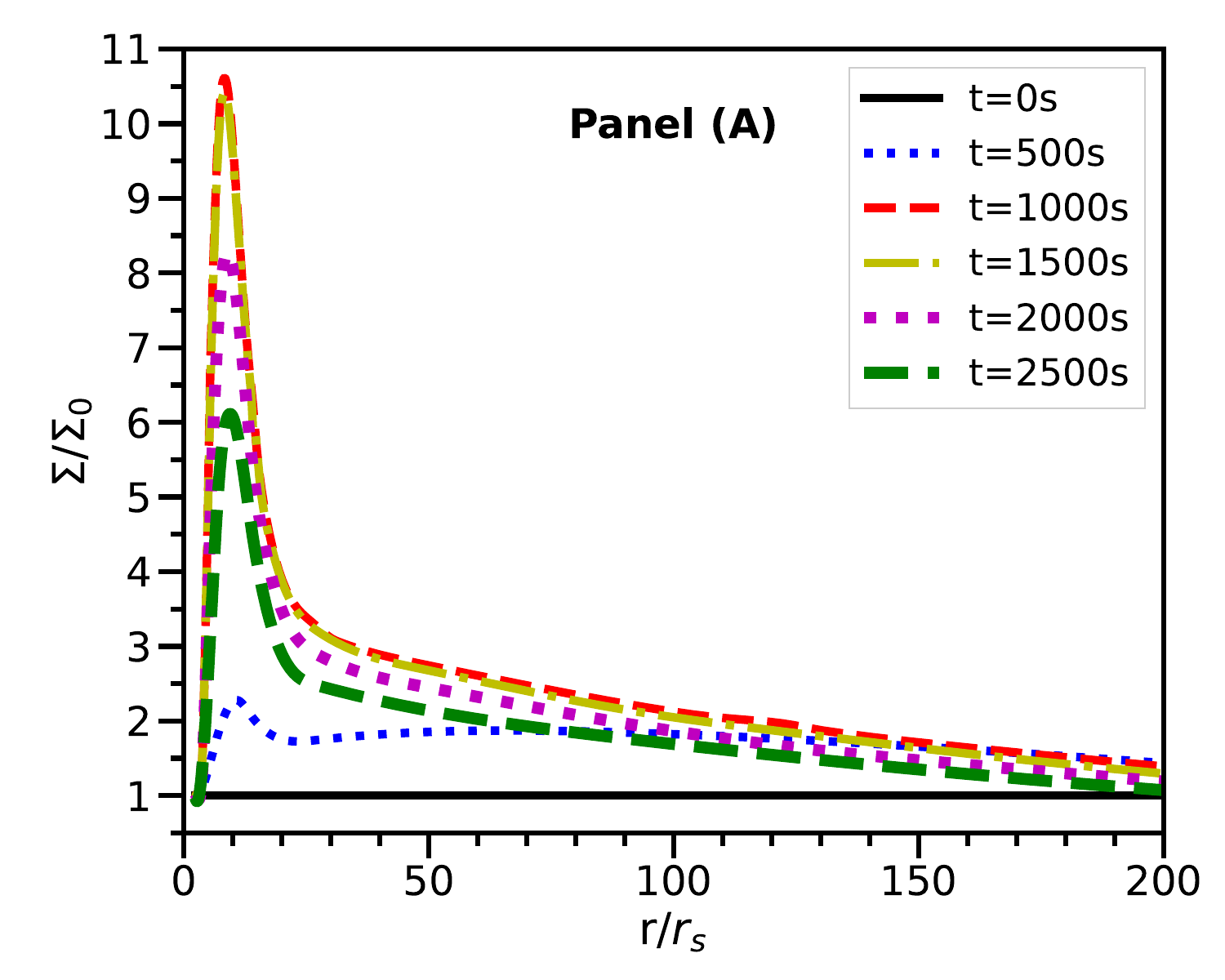}}}
    \scalebox{0.3}[0.3]{\rotatebox{0}{\includegraphics{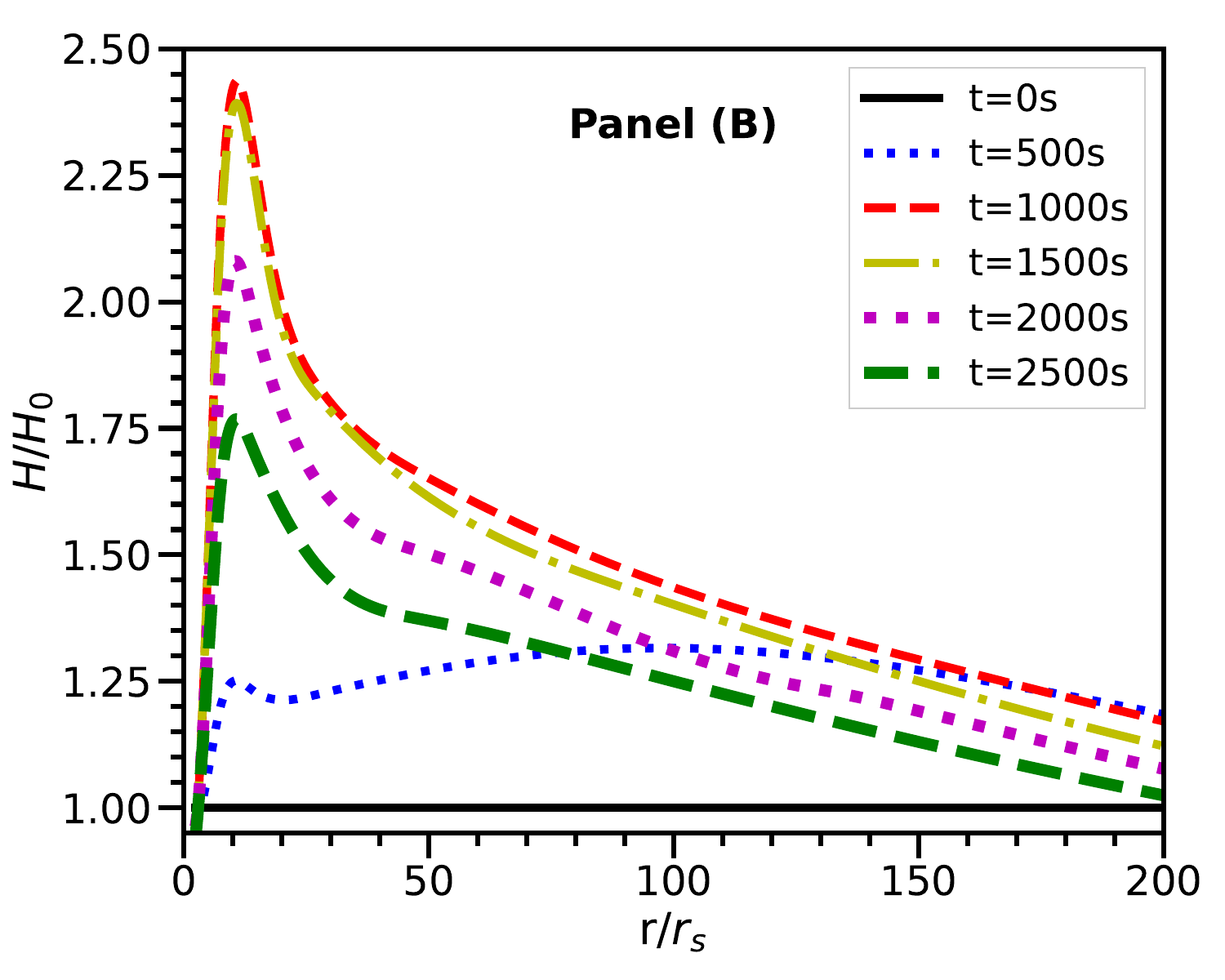}}}\\
    \scalebox{0.3}[0.3]{\rotatebox{0}{\includegraphics{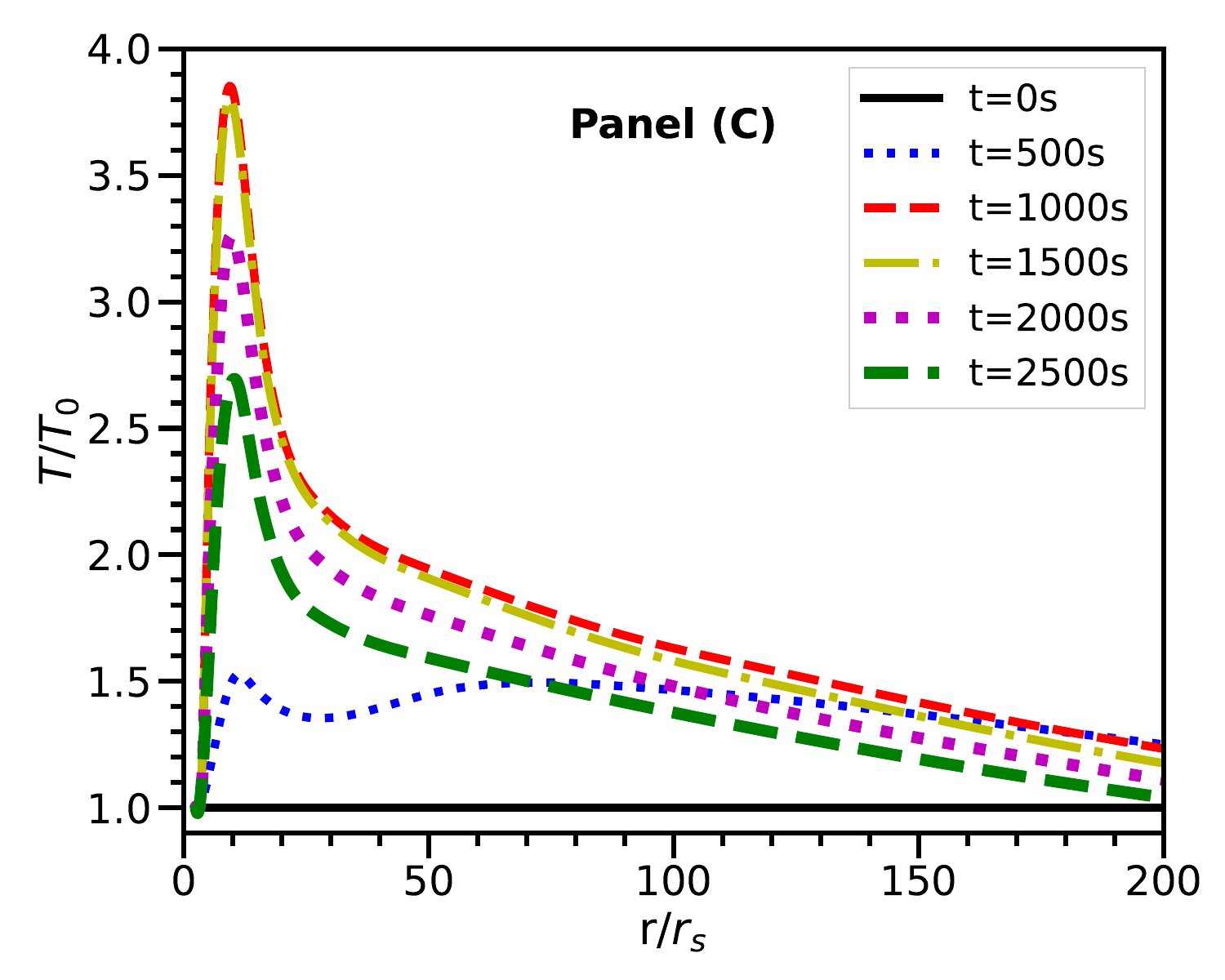}}}
    \scalebox{0.3}[0.3]{\rotatebox{0}{\includegraphics{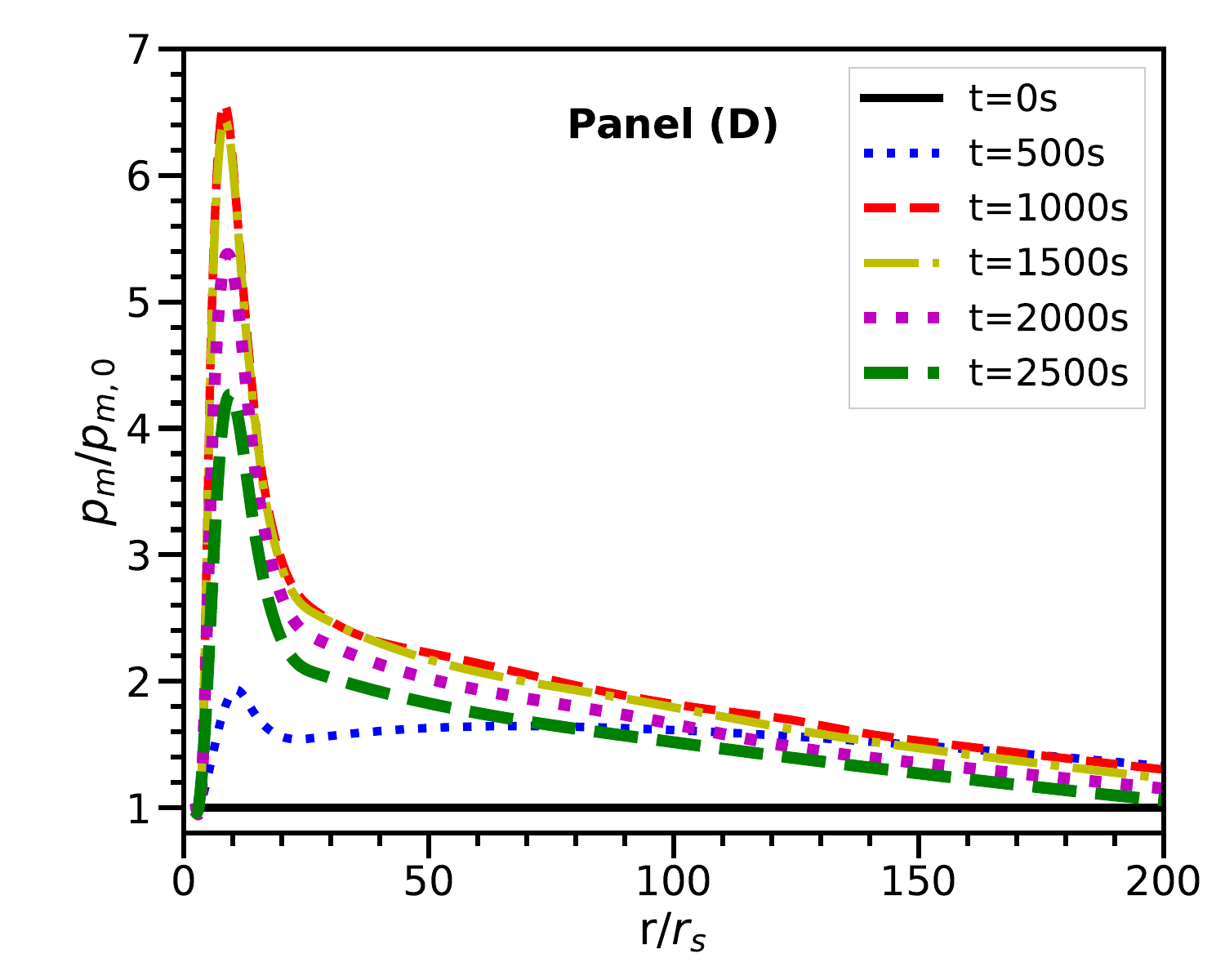}}}\\
    \scalebox{0.3}[0.3]{\rotatebox{0}{\includegraphics{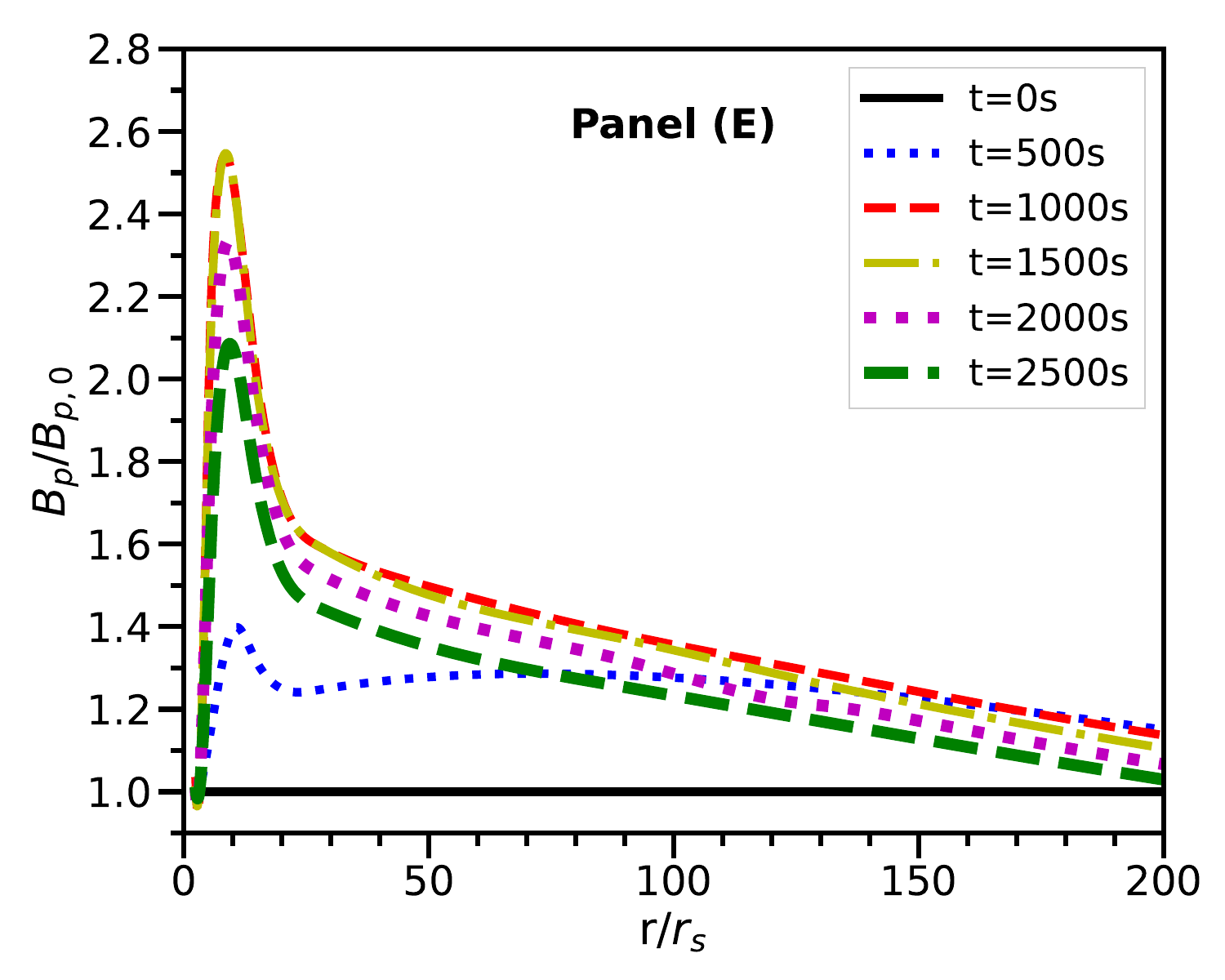}}}
    \scalebox{0.3}[0.3]{\rotatebox{0}{\includegraphics{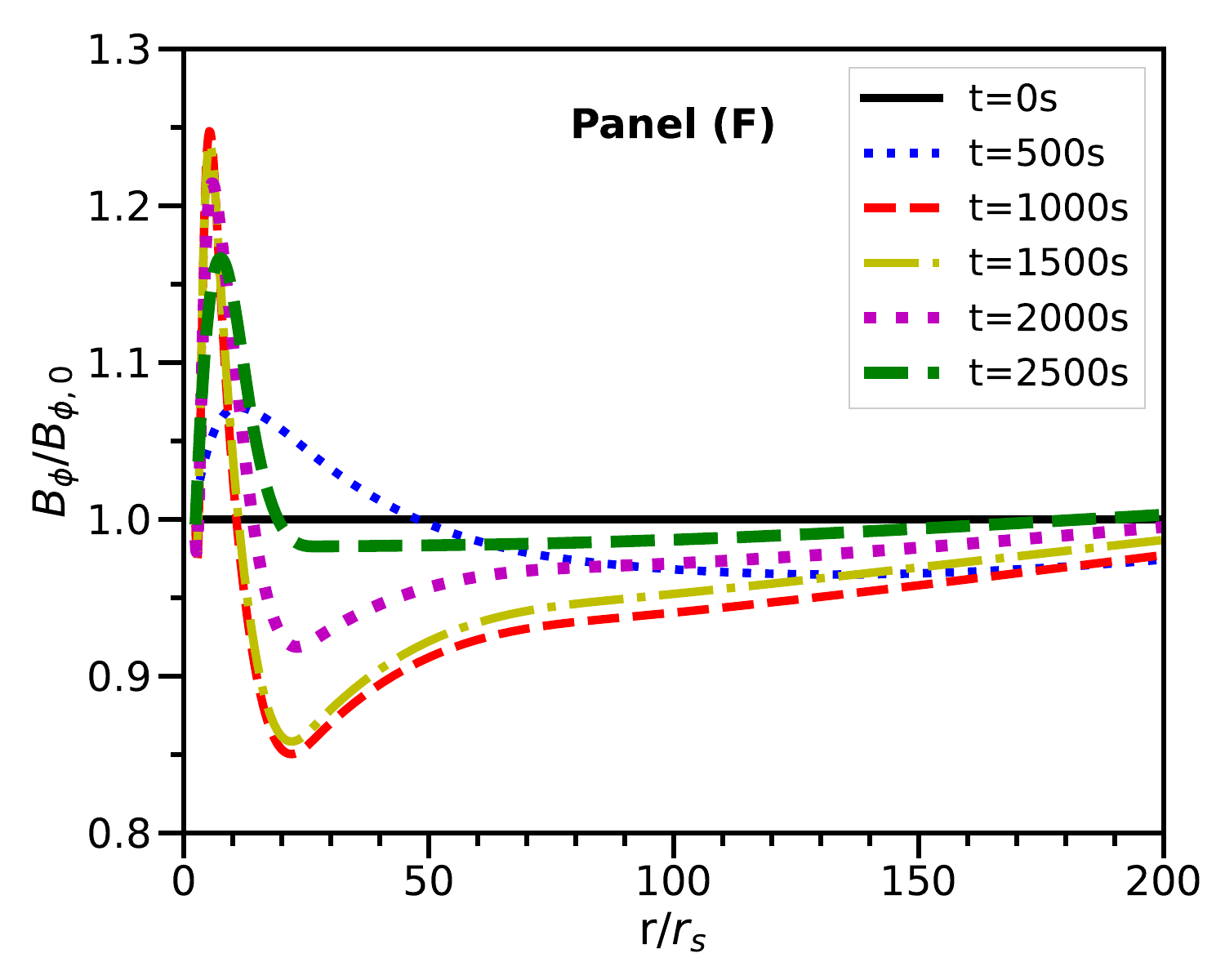}}}
    \ \centering \caption{Evolutions of the disc structures in model M4 with $\beta_{1,0}=100$ and $\mu=1.5$. Panels (A)$-$(D) show the ratio of the surface density, half-height, temperature, and magnetic pressure to their initial values at different time, respectively. Panels (E) and (F) show the ratio of the poloidal and toroidal fields to their initial values at different time, respectively.}
    \label{Fig5}
\end{figure*}

\begin{figure}
    \scalebox{0.35}[0.35]{\rotatebox{0}{\includegraphics{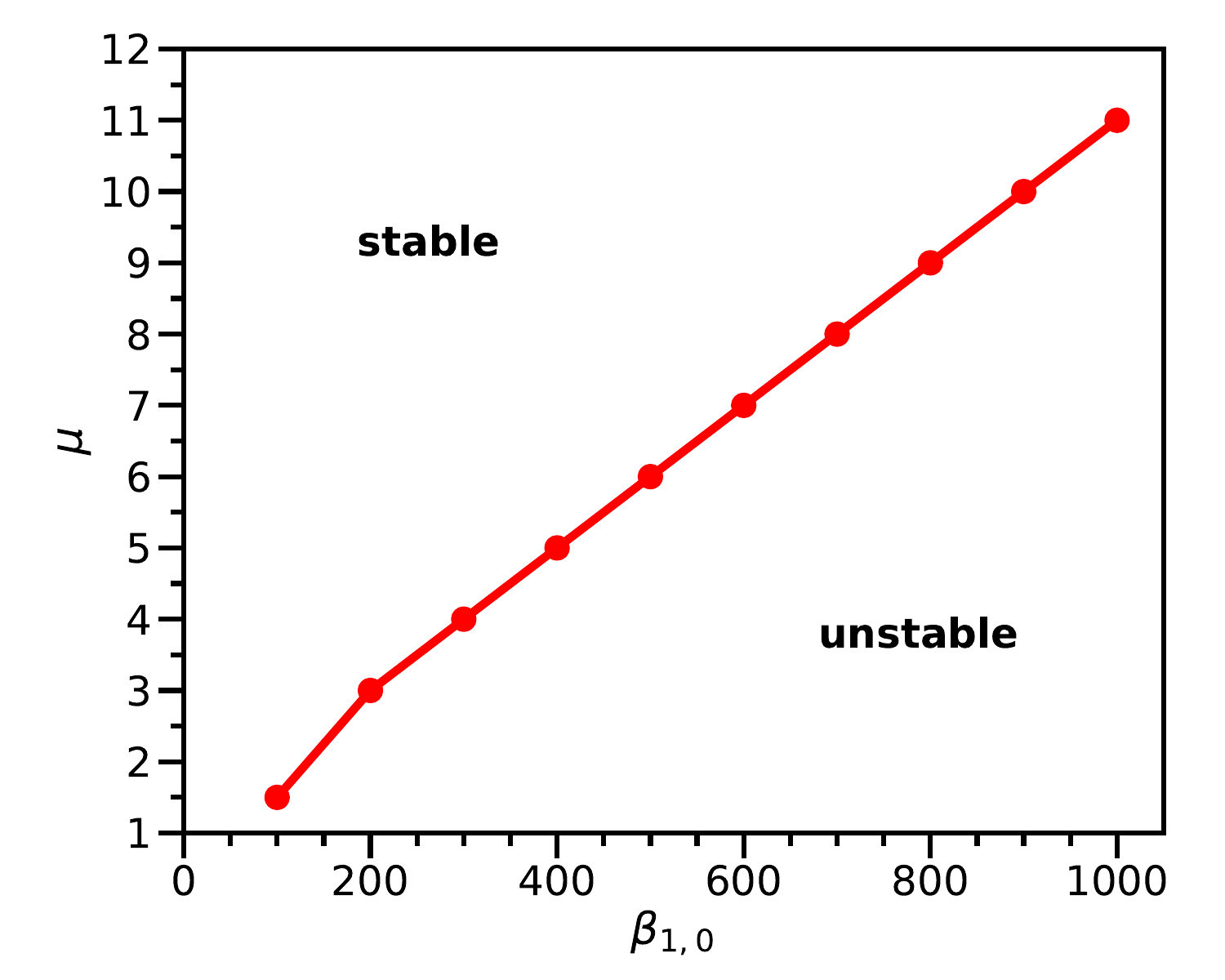}}}
    \ \centering \caption{Boundary between stable and unstable areas in the $\mu$--$\beta_{1,0}$ paramater space for the $B_{\phi,0}=0.1 B_{p,0}$ case. In the $B_{\phi,0}=10B_{p,0}$ case, the $\mu$ value on the boundary is higher 100 times than that in the $B_{\phi,0}=0.1 B_{p,0}$ case.}
    \label{Fig6}
\end{figure}

In this work, we calculate the global time-dependent evolution of a thin accretion disc with magnetic fields to investigate the effects of magnetically driven winds. The magnetically driven winds have two important effects on the thin disc. First, they facilitate the angular momentum transfer, which is related to the strength of magnetic fields. Secondly, these winds contribute to the cooling of the disc, which is related to the mass outflow rate of winds. In our models, the mass outflow rate is proportional to a mass loading parameter ($\mu$). When $\mu$ is small, the mass outflow rate of winds is low, and although the winds play an important role in transferring angular momentum, they cannot be helpful to eliminate the thermal instability. In this case, due to the transfer of angular momentum, the limit-cycle period is shortened. When the initial magnetic field strength is one percent of the sum of gas and radiation pressure, the limit-cycle period is shortened by about 20\% compared to the case without magnetic fields. When $\mu$ is large, the mass outflow rate of winds is high and then the winds carry away enough energy from the disc to suppress the outburst driven by the thermal instability. Therefore, the thin disc becomes thermally stable.

The mass loading parameter $\mu$ is an important parameter to determine whether the thin disc is thermally stable or not, while it is taken as a free parameter in our models. $\mu$ is proportional to $\rho v_{\rm p}/B_{\rm p}$ along magnetic field lines from the definition in \cite{Spruit1996}, and therefore it depends on local gas and field properties. In theory, the value of $\mu$ can be obtained based on the cold approximation of the Weber--Davis model \citep{Weber1967, Spruit1996}, where the thermal pressure is neglected. When models deviate from the cold approximation, what is the exact value of $\mu$ is an open issue. Unfortunately, numerical simulations have not given the range of $\mu$. Under the cold approximation of the Weber--Davis model, the magnetic torque exerted by magnetically driven winds on the thin disc can be written as $T_{\rm m}=3r B_{\rm p}^2\mu(1+\mu^{-2/3})/4\pi$ (e.g. \citealt{Cao2013}). Combining with Equation \ref{Tm}, we have ${B_{\phi}}/{B_{p}}=\frac{3}{2}\mu(1+\mu^{-2/3})$. This is used to determine $\mu$ based on $B_{\phi}$ and $B_{\rm p}$ in \citet{Pan2021}, and then $\mu=2.9\times10^{-4}$ for $B_{\phi}=0.1B_{\rm p}$ while $\mu=5$ for $B_{\phi}=10B_{\rm p}$. However, these values of $\mu$ are much smaller than the critical value required for the thermal stability of the disc, as determined in our calculations. There are probably factors to affect the loading of material on the magnetic field lines. Above the disc surface, it is often suggested that there is a corona. The corona is helpful to load material along the magnetic lines \citep{Cao1994}. Moreover, when the disc expands, the vertical velocity at the disc surface is helpful to push the material along the magnetic lines. Furthermore, as shown in the panel (A) and panel (B) of Fig. \ref{Fig5}, the disc surface density increases faster compared to the disc height, suggesting an increase in density within the disc, so more material can be pushed along the magnetic lines. At the state of disc expanding, the $\mu$ value is expected to be larger than that given in the cold Weber--Davis model. However, the exact value of $\mu$ is unknown now, which is really worth using MHD simulations to further investigate this issue.

Fig. \ref{Fig5} [panel (B)] shows that a thermally stable disc experiences the processes of expanding and shrinking. In our simulations, the disc does not restore to its initial state after the disc shrinks, although it is expected that the disc should restore to its initial state. This is because strong winds always exist, whether the disc expands or shrinks. In fact, the mass loading parameter $\mu$ should be a function of radius and time. It increases when the disc expands, while it decreases when the disc shrinks. In the this work, we set it constant with radius and time. This means that when the disc shrinks the strong winds still exist and then the disc in our simulations cannot be restored to its initial state.

\section{ACKNOWLEDGMENTS}
XH is supported in part by the Natural Science Foundation of China (grant 11973018) and Chongqing Natural Science
Foundation (grant CSTB2023NSCQ-MSX0093). XL is supported by the Natural Science Foundation of Fujian Province of China(No. 2023J01008). SL is supported in part by the Natural Science Foundation of China (grants 12273089).

\section{DATA AVAILABILITY}
The data underlying this article will be shared on reasonable request to the corresponding author.

\bibliographystyle{mnras}
\bibliography{ref}

\end{document}